\documentclass[twocolumn,english,aps,prb,twocolum,superscriptaddress,bibnotes,amsmath,amssymb,floatfix]{revtex4-1}
\usepackage[colorlinks=true,citecolor=blue,linkcolor=magenta]{hyperref}

\usepackage{changes}
\usepackage{soul}
\usepackage[utf8]{inputenc}
\usepackage[english]{babel}
\usepackage{amsmath,amsfonts,amssymb}
\usepackage[T1]{fontenc}
\usepackage{url}

\usepackage{amsmath}
\usepackage{amsfonts}
\usepackage{amssymb}
\usepackage{setspace}
\usepackage[version=3]{mhchem}
\usepackage{lmodern}
\usepackage{soul}
\usepackage{changes}
\usepackage{epstopdf}
\usepackage{graphicx}

\begin{document}
\title{Ultrafast optical circuit switching for data centers\\using integrated soliton microcombs}

\author{A. S. Raja}
\thanks{These authors contributed equally to this work.}
\affiliation{Institute of Physics, Swiss Federal Institute of Technology Lausanne (EPFL), CH-1015 Lausanne, Switzerland}

\author{S. Lange}
\thanks{These authors contributed equally to this work.}
\affiliation{Microsoft Research, 21 Station Road, Cambridge, CB1 2FB, U.K.}

\author{M. Karpov}
\thanks{These authors contributed equally to this work.}
\affiliation{Institute of Physics, Swiss Federal Institute of Technology Lausanne (EPFL), CH-1015 Lausanne, Switzerland}

\author{K. Shi}
\thanks{These authors contributed equally to this work.}
\affiliation{Microsoft Research, 21 Station Road, Cambridge, CB1 2FB, U.K.}

\author{X. Fu}
\affiliation{Institute of Physics, Swiss Federal Institute of Technology Lausanne (EPFL), CH-1015 Lausanne, Switzerland}

\author{R. Behrendt}
\affiliation{Microsoft Research, 21 Station Road, Cambridge, CB1 2FB, U.K.}

\author{D. Cletheroe}
\affiliation{Microsoft Research, 21 Station Road, Cambridge, CB1 2FB, U.K.}

\author{A. Lukashchuk}
\affiliation{Institute of Physics, Swiss Federal Institute of Technology Lausanne (EPFL), CH-1015 Lausanne, Switzerland}

\author{I. Haller}
\affiliation{Microsoft Research, 21 Station Road, Cambridge, CB1 2FB, U.K.}

\author{F. Karinou}
\affiliation{Microsoft Research, 21 Station Road, Cambridge, CB1 2FB, U.K.}

\author{B. Thomsen}
\affiliation{Microsoft Research, 21 Station Road, Cambridge, CB1 2FB, U.K.}

\author{K. Jozwik}
\affiliation{Microsoft Research, 21 Station Road, Cambridge, CB1 2FB, U.K.}

\author{J. Liu}
\affiliation{Institute of Physics, Swiss Federal Institute of Technology Lausanne (EPFL), CH-1015 Lausanne, Switzerland}

\author{P. Costa}
\affiliation{Microsoft Research, 21 Station Road, Cambridge, CB1 2FB, U.K.}

\author{T. J. Kippenberg}
\email[]{tobias.kippenberg@epfl.ch}
\affiliation{Institute of Physics, Swiss Federal Institute of Technology Lausanne (EPFL), CH-1015 Lausanne, Switzerland}

\author{H. Ballani}
\email[]{hitesh.ballani@microsoft.com}
\affiliation{Microsoft Research, 21 Station Road, Cambridge, CB1 2FB, U.K.}

\date{\today}
\begin{abstract}
\textbf{Networks inside current data centers comprise a hierarchy of  power-hungry electronic packet switches interconnected via optical fibers and transceivers. 
As the scaling of such electrically-switched networks approaches a plateau, a power-efficient solution is to implement a flat network with optical circuit switching (OCS),  without electronic switches and a reduced number of transceivers due to direct links among servers. One of the promising ways of implementing OCS is by using tunable lasers and arrayed waveguide grating routers. Such an OCS-network can offer high bandwidth and low network latency, and the possibility of photonic integration results in an energy-efficient, compact, and scalable photonic data center network.  To support dynamic data center workloads efficiently, it is critical to switch between wavelengths in sub nanoseconds (ns). Here we demonstrate ultrafast photonic circuit switching based on a microcomb. Using a photonic integrated $\mathbf {Si_3N_4}$ microcomb in conjunction with semiconductor optical amplifiers (SOAs), sub ns (< 500 ps) switching of more than 20 carriers is achieved.  Moreover, the 25-Gbps non-return to zero (NRZ) and 50-Gbps four-level pulse amplitude modulation (PAM-4) burst mode transmission systems are shown. Further, on-chip Indium phosphide (InP) based SOAs and arrayed waveguide grating (AWG) are used to show sub-ns switching along with 25-Gbps NRZ burst mode transmission providing a path toward a more scalable and energy-efficient wavelength-switched network for future data centers. }
\end{abstract}
\maketitle
\noindent\textbf{\noindent}
\def \DWRep {\Delta \omega_{\rm rep}}
\def \DFRep {\Delta f_{\rm rep}}
There is a massive growth in data center network traffic due to emerging applications such as AI, server-less computing, resource disaggregation, multimedia streaming, and big data\cite{Randy2009}. These applications are evolving quite drastically, requiring data centers networks with low-latency, high-bandwidth, and high scalability. Current data center networks comprise a multi-tier interconnection of electronic switches and optical transceivers that consume a lot of power, particularly at high network bandwidths. Further, continued bandwidth scaling of the electrical switches poses a significant challenge\cite{ballani2018, Cheng:18}. 

Optical circuit switching (OCS) has been proposed as an alternative technology to overcome these challenges; it can provide high bandwidth and low network latency (due to no network buffers), and by allowing for a flat data center topology, the network can be more energy-efficient as fewer electrical switches and transceivers are required\cite{barry2018}.  In OCS the servers are optically interconnected directly without needing the opto-electronic-opto (O-E-O) conversion. Micro-electro-mechanical (MEMS) based OCS systems are commercially available with the possibility of high port counts ($\sim$ 1000 ports)\cite{han2014,kim2003}. However, MEMS-based OCS architectures suffer from slow switching time ($\sim$ ms), affecting efficient utilization of network resources, and require extreme fabrication precision. In contrast, modern data centers require fast switching ($<$ ns) to efficiently support emerging hardware-accelerated applications; for applications like key-value stores, > 97\% of the packets have a size of 576 bytes or less\cite{clark2018}. Many photonic chip-based techniques have been used to demonstrate the fast optical switching (ns) such as semiconductor optical amplifiers (SOAs) \cite{tanaka2009}, electro-absorption modulators (EAMs) \cite{segawa2013}, and Mach–Zehnder interferometers (MZIs)\cite{dupuis2019,dupuis2020}. The arrayed waveguide grating routers (AWGRs) \cite{yin2013,xiao_2019}, in conjunction with a tunable laser (TL), is a promising optical circuit switching technology as the core of the network is passive which benefits both fault-tolerance and future scaling. Recent studies have shown that the tuning of the laser plays a major role in switching speed, and it has been indeed improved from 80 ns to 5 ns by optimizing the designs \cite{simsarian2006,funnell2017}. Moreover, sub-ns switching system is demonstrated recently by ``disaggregating'' or separating the lasing from the actual tuning\cite{shi2019, sigcomm2020}. This has been achieved, for example, by using a discrete bank of lasers as a multi-wavelength source and a wavelength selector containing photonic integrated arrayed waveguide grating (AWG) and SOAs. The SOAs are preferred as switching elements due to fast and lossless switching, chip-scale integration, and high extinction ratio \cite{tanaka2009}. 

\begin{figure*}[ht]
\centering
\includegraphics[width=1.\textwidth]{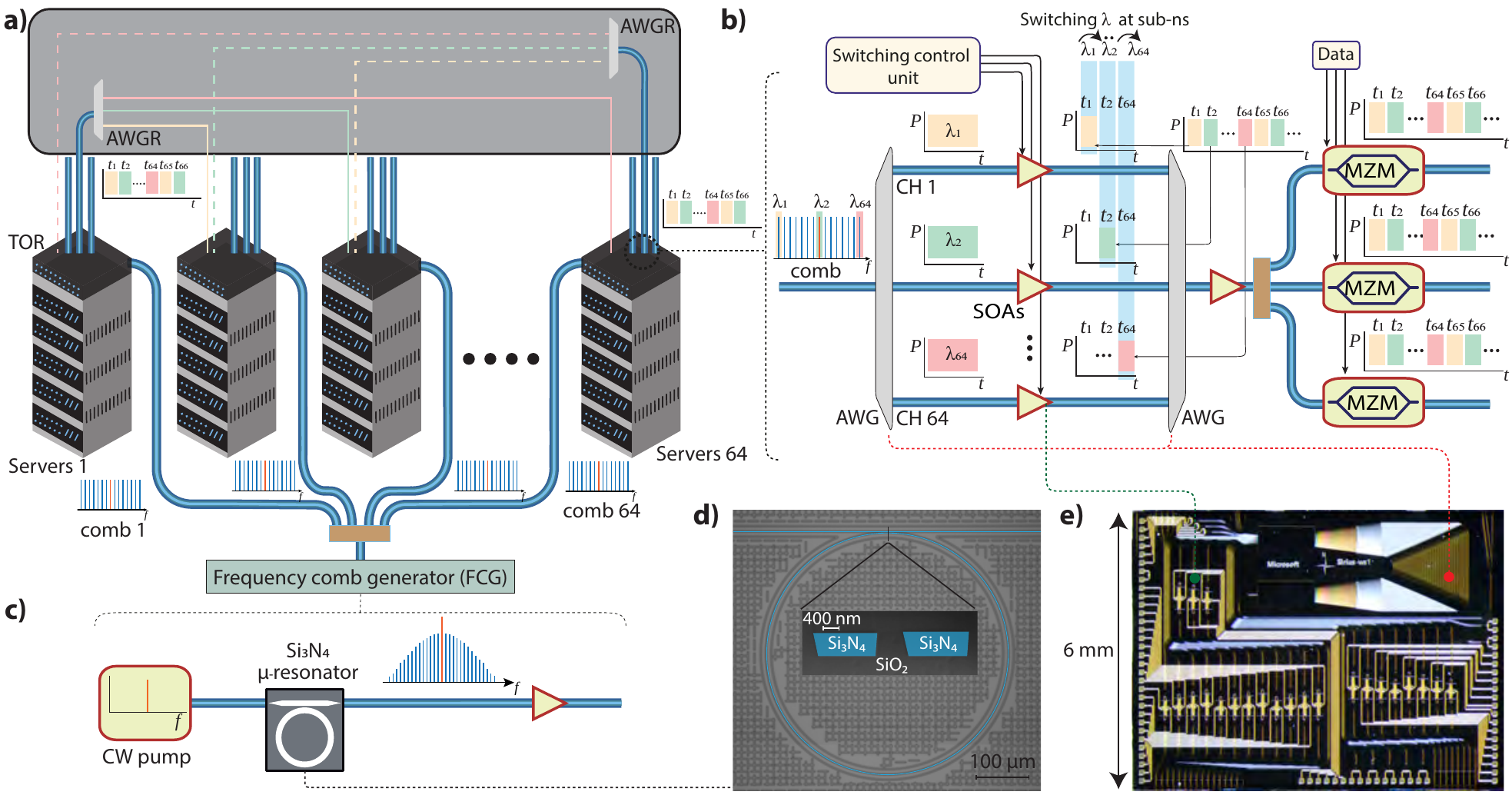}
\caption{\textbf{Concept of optical circuit switching (OCS)  using a photonic chip-based $\mathbf{Si_3N_4}$ soliton microcomb and semiconductor optical amplifiers (SOAs)}. \textbf{a)} Interconnection of 64 servers via an arrayed waveguide grating router (AWGR) for implementing fast OCS. In this model, distinct wavelengths are assigned to each server at each time slot. At each receiver, a 10 to 20 nanoseconds (ns) time slot is assigned for each server on a round-robin basis for data transmission. The switching module is placed on the top of rack (TOR) switch.  A single comb source which is post amplified via cascaded amplifiers to attain high optical power per line can be distributed among many servers. The 64 individual combs (comb1, ..., comb 64) split from a central frequency comb generator (FCG) can be distributed across 64 different servers as a multiwavelength laser, making this architecture more power-efficient and flexible. The multiple data-carrying optical carriers are routed using a passive AWGR to the assigned server. \textbf{b)} Each comb channel is transmitted to SOAs after de-multiplexing, where a control signal (turning on/off current) is applied to switch between the comb channels at sub-ns. The comb channels (10-ns slots) are encoded with data using Mach-Zehnder modulators (MZMs) and transmitted to the relevant servers . The multiple MZMs shown here indicate that this architecture can be scaled further to establish links between more servers. \textbf{c)} The multi-wavelength source based on the chip-scale soliton microcomb is generated by pumping with a single laser.  Microscope images of a $\mathrm{Si_3N_4}$ microresonator \textbf{(d)} and a photonic chip \textbf{(e)}  containing an AWG and SOAs. The inset in \textbf{(d)} shows a false color SEM image of a  $\mathrm{Si_3N_4}$ microresonator's coupling section. }
\label{Fig:figure1}
\end{figure*}

Here we propose and demonstrate in a proof-of-concept  experiment a disaggregated tunable transceiver that uses photonic chip-based soliton microcombs as a multi-wavelength source, which are generated in high-quality ($Q$)  microresonators exhibiting third-order nonlinearity ($\chi^3$)  and anomalous dispersion\cite{kippenberg2018,Herr20114}. The soliton microcombs have been used in many system-level applications, for example, distance ranging (LiDAR) \cite{trocha2018, suh2018, riemensberger2020},  microwave photonics \cite{TorresCompany:14,Wu:18, suh2018b, liu2020}, optical coherence tomography (OCT) \cite{ji2019, paul2019}, and coherent communication \cite{marin2017, fulop2018, corcoran:2020}. The remarkable improvements in the losses of high-confinement $\mathrm{Si_3N_4}$ integrated waveguides, a widely used platform for integrated photonics \cite{Moss:13,Blumenthal:18}, allows the generation of solitons using the on-chip laser \cite{stern2018,pavlov2018, raja2019}. The broadband-bandwidth operation across C- and L-bands, precise spacing control to match ITU frequency grid (50/100 GHz), low power consumption, and wafer-scale scalability make the soliton microcomb an attractive solution for an ultra-fast wavelength switched system in comparison with a bank of lasers \cite{marin2017}. Further, no extra guard band is required for frequency stabilization or to align each comb channel to the frequency grid. 

\begin{figure*}[ht]
\centering
\includegraphics[width=1.\textwidth]{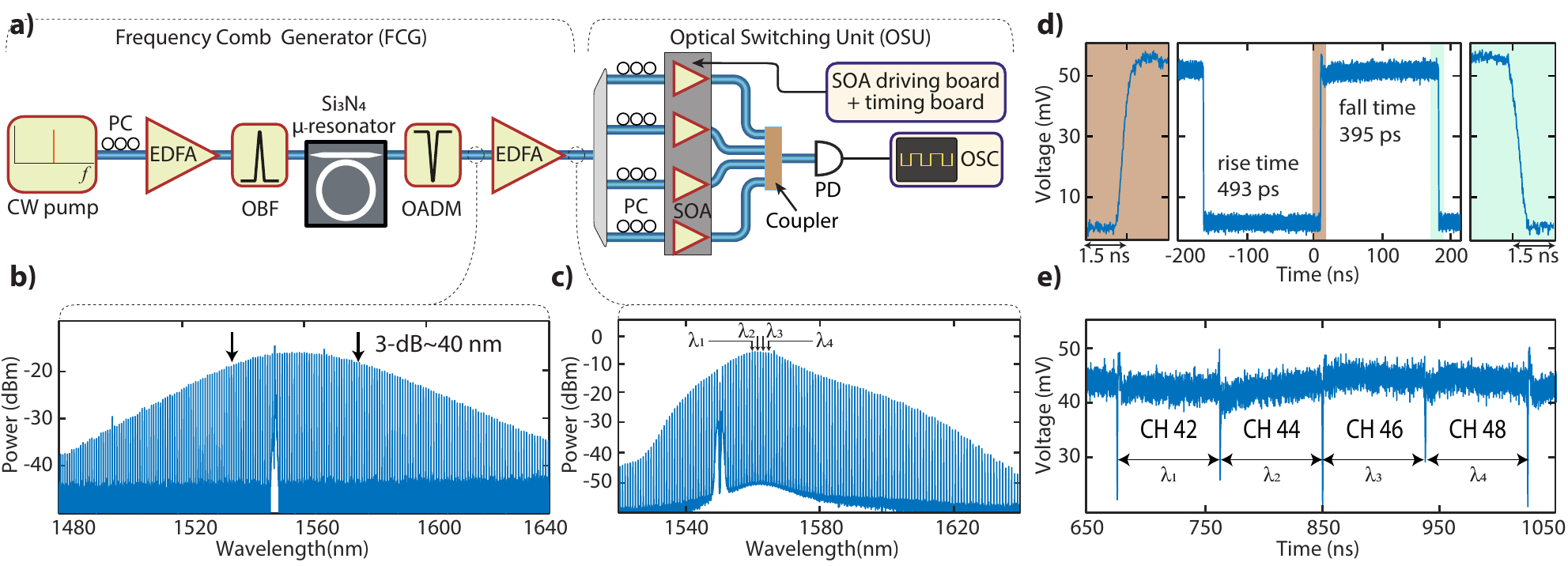}
\caption{\textbf{Experimental demonstration of sub-ns optical circuit switching (OCS) of single and four different microcomb channels.} \textbf{a)} A packaged $\mathrm{Si_3N_4}$ microresonator is pumped using a continuous wave (CW) pump amplified via an erbium-doped fiber amplifier (EDFA) to generate a soliton microcomb. A soliton state is initiated via forward tuning. After filtering the pump using an optical add-drop multiplexer (OADM), a low-noise amplifier is used to further amplify the soliton comb. The post-amplified comb is de-multiplexed using a 48-channels 100G  AWG, and the individual comb channels are routed to semiconductor amplifiers (SOAs) which are controlled via a custom-designed switching board. The output signal is detected using a photodiode (PD) and recorded on an oscilloscope (OSC). PC, polarization controller; OBF, (narrow) optical bandpass filter. \textbf{b)} A single soliton spectrum with 3-dB bandwidth around 40 nm  and no post amplification. \textbf{c)} Post-amplified soliton  spectrum having maximum power up to -4-dBm  in comb lines close to pump line (1550 nm). \textbf{d)} The 10\%$\times$90\% rise and fall time of 493 ps and 395 ps respectively for single comb channel (CH 37: 1554.9 nm). The left and right insets show the zoomed-in view of rising and falling signals. \textbf{e)} The four different microcomb channels simultaneously switching at sub-ns time scale with wavelength separation $\sim$ 4.9 nm. A guard zone of $ \sim$ 2.56 ns is used to allow smooth switching between two adjacent  channels. CH-42: 1558.988 nm, CH-44: 1560.613 nm, CH-46: 1562.238 nm, and CH-48: 1563.863 nm.    }
\label{Fig:figure2}
\end{figure*}

 We use  a $\mathrm {Si_3N_4}$ based soliton microcomb as a multi-wavelength source to show ultrafast ($<$ 550 ps) optical wavelength switching. In a proof-of-concept experiment, we demonstrate switching of $>$ 20 individual comb channels in a sub-ns time scale ($<$ 550 ps) using discrete SOAs. Further, 25-GBd (baud-rate) no-return to zero on-off keying (NRZ) and four-level pulse amplitude modulation (PAM-4) burst mode transmission  systems along with ultrafast switching are shown. Then, a more compact switching system consisting of PIC-based AWG and SOAs is implemented to show sub-ns switching and 25-GBd NRZ burst mode transmission, indicating the potential utilization of such a miniaturized system to mitigate power and scaling issues.
A key advantage of the disaggregated transceiver design is that the multi-wavelength comb source can be shared across many (e.g.  64) servers in parallel instead of using 64 separate comb sources, using a split-and-amplify architecture (figure \ref{Fig:figure1}a). The source can thus be treated as a shared infrastructure element such as the power source in today's data centers. This allows for an appealing division of functionality since  the power consumed by the comb source is amortized as the power efficiency of the end-to-end system converges to the efficiency of the amplifiers while allowing for a high-quality and stable light source that can be rapidly wavelength-tuned (cf. discussion on power analysis and Supplementary information (SI)).

Figure \ref{Fig:figure1} shows the OCS architecture containing the soliton microcomb, SOAs, AWGs in the switch and AWGRs to route the data across many servers (cf. Method).  This architecture allows further  parallelization of different resources by sharing the soliton across many servers and modulators for power efficiency and parallel data transmission, respectively (figure \ref{Fig:figure1}a \& b ).  The multi-wavelength source is generated by pumping a packaged $\mathrm{Si_3N_4}$ microresonator fabricated using the photonic Damascene reflow \cite{Pfeiffer2018,Liu2018a} process enabling a mean intrinsic $Q$-factor ($Q_0$) of > 15 million. Initially, a multi-soliton is initiated by performing a scan over resonance (forward tuning) and then a single soliton is generated via backwards switching \cite{guo2017} (Figure \ref{Fig:figure2}b).  The soliton is amplified using a low-noise and compact EDFA resulting in a comb with a maximum power of up to -4 dBm and an optical signal-to-noise ratio (OSNR) of > 34 dB. The post-amplified soliton as shown in figure  \ref{Fig:figure2}c  is de-multiplexed  using a 100G spaced 48 channels AWG ($\sim$ 1525 $\mathrm{nm}$-1564 $\mathrm{nm}$) providing 30-dB isolation.  The individual comb channels are initially switched using discrete SOAs with a small-signal gain $\sim$ 11-13 dB at 1550 nm. Figure \ref{Fig:figure2}d shows a 10\% $\times$ 90\% rise and fall times of  493 ps and 395 ps, respectively, for a single microcomb carrier centered at 1555 nm (CH 37 of AWG) when applying a current of 120 mA to operate the SOA.
Similarly, more than 20 comb channels (1540 $\mathrm{nm}$ - 1564 $\mathrm{nm}$) in C-band are tested individually to show sub-ns switching (cf. SI). Even though not tested due to the unavailability of an L-band AWG, the current results indicate that more than 40 comb channels in the L-band could be used as well. 
\begin{figure*}[ht]
\centering
\includegraphics[width=1.\textwidth]{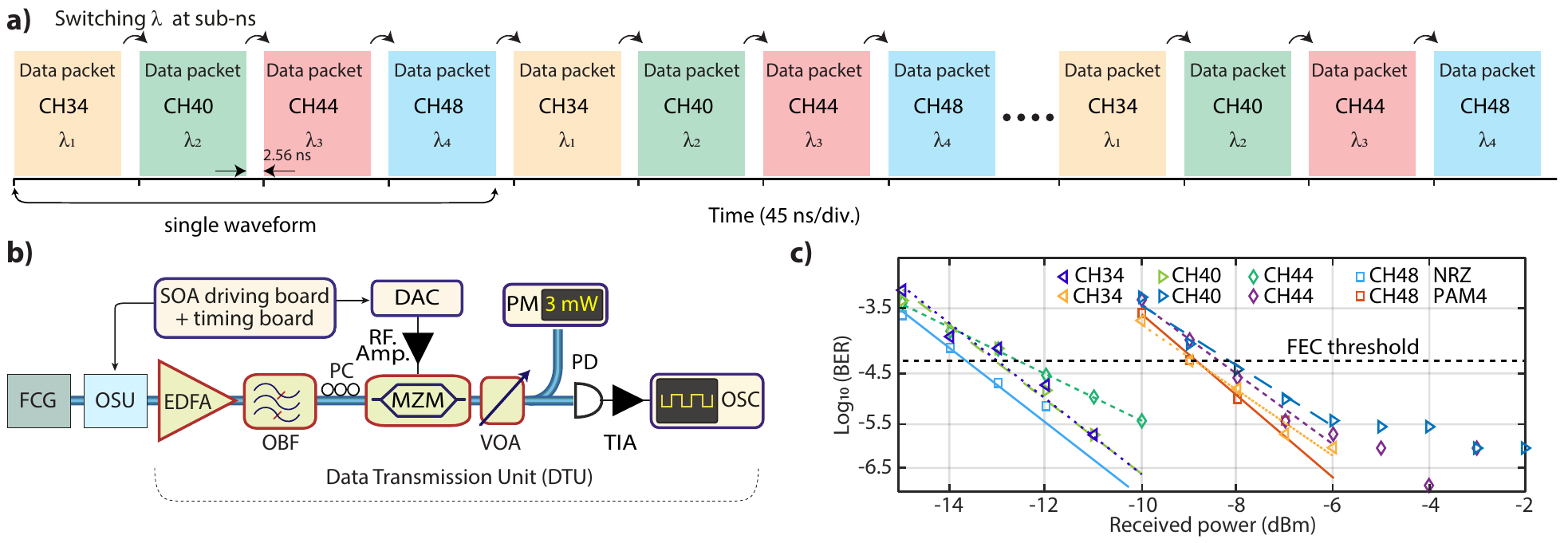}
\caption{\textbf{Experimental demonstration of burst mode NRZ and PAM-4 transmission using discrete SOAs while switching.} \textbf{a)}  A stream of multiple data packets showing data transmission along with sequential switching between the four comb channels. A single burst waveform sequence consists of header, payload, and guard zone containing 32, 1024, and 64 symbols respectively. \textbf{b)} The signal after the frequency comb generator (FCG) and the optical switching unit (OSU) is amplified using a compact EDFA to compensate for losses of the 20-GHz Mach-Zehnder modulator (MZM).  Then, it is filtered out using a wide-band optical bandpass filter (OBF) ($\sim$ 20 nm) to reject amplified spontaneous emission (ASE) noise from the SOAs and the EDFA. The data is encoded on the modulator using an arbitrary waveform generator (DAC). A fast photodiode (PD) is used to detect the signal. The electrical signal is amplified using a trans-impedance amplifier (TIA) and finally captured by an OSC. \textbf{c)}  The bit error ratio (BER) of four different comb channels while switching between them, using different modulation formats non-return to zero (NRZ) and four-level pulse amplitude modulation (PAM-4). A performance below forward error correction (FEC) threshold is achieved for NRZ and PAM-4 for a received optical power of $\sim$ -12 dBm and $\sim$ -8 dBm, respectively.  CH-34: 1552.524 nm, CH-40: 1557.363 nm, CH-44: 1560.606 nm, and CH-48: 1563.863 nm.}
\label{Fig:figure3}
\end{figure*}

For a proof-of-concept system-level demonstration, fast switching within a four different comb channels is performed.  Figure \ref{Fig:figure2}e shows 4 different comb channels switching at ns-timescale with a separation of $\sim$ 5.6 $\mathrm{nm}$. A guard zone of 2.56 ns is used to allow a smooth switching between adjacent comb channels while an external reference clock (timing board) aligns the different switching signals.  Distinct currents are applied to the SOAs to achieve a constant output power  and to compensate for the non-uniform comb power per line or SOAs gain (figure \ref{Fig:figure2}e). While the sub-ns switching of four channels with a maximum 20 nm separation has been demonstrated, the maximum channel separation is mainly limited by the optical band-pass filter (cf. SI). 

In the following experiment, we show 25 Gbps (NRZ) and 50 Gbps  (PAM-4)  burst mode data transmission while switching between four comb channels using the setup shown in figure \ref{Fig:figure3}b. The four optical carriers after switching are further amplified to overcome the insertion loss ($ \mathrm{\sim 7-dB}$) of the 20-GHz Mach-Zehnder modulator (MZM) which is operating at the quadrature point. In addition to eliminating comb channels in the next order FSR of the AWG, the OBF is utilized to suppress the out-of-band SOA and EDFA amplified spontaneous emission (ASE) noise. The burst mode sequence at 25 GBaud symbol rate, generated by the arbitrary waveform generator is applied to the MZM with a random sequence of $2^{15}$- and $2^{16}$-bits for NRZ and PAM-4 respectively. The electrical waveform is amplified using a trans-impedance amplifier (TIA) after detecting it on a fast photodiode having 50-GHz bandwidth. The amplified waveform is acquired using a real-time oscilloscope with 160 GSamples/s sampling rate. The digital signal processing (DSP), explained in detail in ref \cite{shi2019}, is performed offline to obtain the bit error ratio (BER). The received optical power (ROP) vs. (BER) of the system, as shown in figure \ref{Fig:figure3}c, is characterized by changing the optical power of incoming waveform via a variable optical attenuator (VOA). A BER of below $5 \times 10^{-5}$, which is the threshold for forward error correction (FEC) in data center transmission systems, is achieved for both NRZ and PAM-4 at a ROP of $\sim$ -12 dBm and $\sim$ -8 dBm respectively. The BER error floor for PAM4 at a ROP of > -6 dBm emerges due to ASE and AWG crosstalk.
\begin{figure*}[ht]
\centering
\includegraphics[width=1.\textwidth]{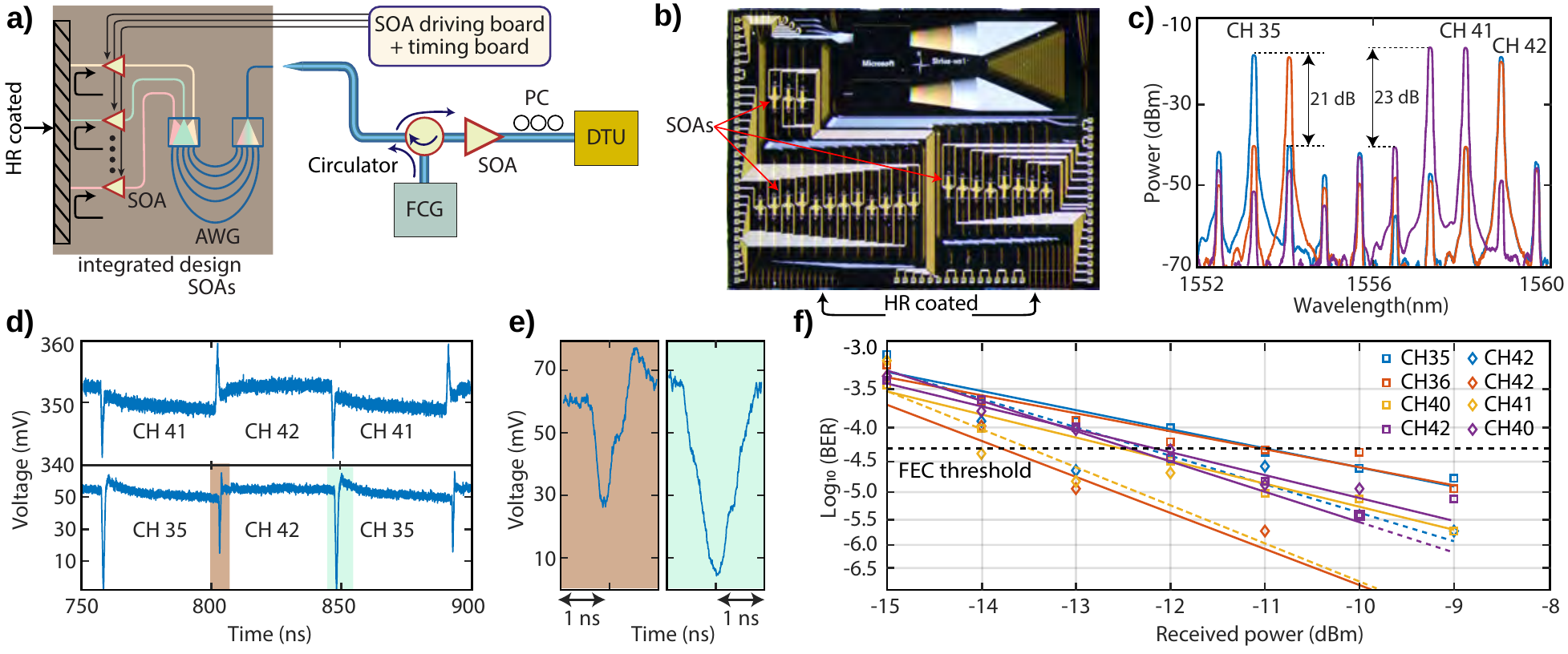}
\caption{\textbf{Sub-ns optical circuit switching (OCS) and data transmission using  on-chip SOAs and on-chip AWG along with soliton microcomb.} \textbf{a)} Schematic of the setup used to perform the OCS and data transmission. The multi-wavelength optical carriers, generated via the frequency comb generator (FCG), are coupled to an InP chip containing an AWG and SOAs via an optical circulator. The coupled optical carriers are aligned to the AWG by changing the temperature of InP chip. The aligned carriers are transmitted to integrated SOAs; if one of them is biased, then the AWG channel (waveguide) connected to that particular SOA is reflected from the high-reflection coated facet of the chip while non-biased SOAs block the light. The reflected-back channels are coupled back via an anti-reflection coated optical fiber for encoding the information using the data transmission unit (DTU).  \textbf{b)} Microscope image of PIC showing the SOAs (red arrows) and AWG. \textbf{c)} Optical spectrum of different comb channels after the InP chip when biasing the on-chip SOAs indicating more than 20 dB isolation with adjacent AWG channels (cross talk). \textbf{d)} The sub-ns wavelength switching between two different comb channels using on-chip SOAs. The overshoot in the switching signal is due to impedance mismatch between the high-speed radio-frequency (RF) probes and the on-chip electrodes. This effect can be minimized by optimizing the drive signal. \textbf{e)} The left and right figures show the zoomed-in view of switching signals between two different comb channels (CH35 and CH 42). \textbf{f)} The bit error ratio (BER) performance of  the 25 GBd NRZ PIC-based switching system for different combinations of two comb channels.}
\label{Fig:figure4}
\end{figure*}

Next, integrated Indium phosphide (InP) chip-based SOAs and AWG are used to show 25-Gbps NRZ burst mode data transmission along with fast OCS using a soliton microcomb.   Figure \ref{Fig:figure4}b shows the photonic integrated circuit (PIC) based wavelength selector PIC with a dimension of 6 mm $\times$ 8 mm.
The reflection of the light from the high-reflection (HR) coated facet allows simultaneous utilization of an AWG with 32$\times$50 GHz separated channels as multiplexer and de-multiplexer. This simplifies the wavelength alignment procedure and reduces the footprint of the device. Nineteen SOAs are connected via integrated waveguides to AWG output channels.  The wavelength alignment of the comb channels to the AWG is performed by changing the temperature of the PIC, resulting in seven comb channels matching with the AWG. This can also be realized by changing the temperature of the $\mathrm{Si_3N_4}$ chip. Figure \ref{Fig:figure4}c shows the optical spectrum of the AWG aligned comb channels indicating a >20 dB isolation with adjacent channels. Initially, the 10\% $\times$ 90\% rising and falling times of PIC is characterized by performing simultaneous switching between two comb channels. The maximum (minimum) experimentally observed switching time is $\sim$ 820 ps ($\sim$ 375 ps). Moreover, the overshoot in the switching signal, as seen in figure \ref{Fig:figure4}d, arises due to impedance mismatch between the SOAs on-chip electrodes and the RF probes \cite{shi2019}. Then,   a burst mode data transmission demonstration with 25-Gbps NRZ modulation is performed. A BER below FEC threshold is obtained when switching between two-channels with different combinations for a ROP $>$ -11 dBm. The PAM-4 burst mode transmission demonstration requires further improvement in the output power of the comb due to low in- and out-coupling in a packaged $\mathrm{Si_3N_4}$. The main reason behind the low coupling efficiency (15 \%) is an additional 2-dB splicing loss between UHNA and SMF-28 fiber, which can be reduced to $<$ 0.2 dB by using state-of-art splicing instruments\cite{yin2019}.  Similarly, the AWG crosstalk and insertion of loss of the PIC can be further improved to enhance the overall performance of this architecture. Nevertheless, the current results show the potential of a soliton microcomb as a suitable  multi-wavelength source for 25 GBd NRZ burst mode transmission along with fast switching.  



Regarding the power consumption, the current multi-wavelength source  consumes a total electrical power of $\sim$ 30 W (cf. SI) providing more than 60 carriers, having an optical power $>$ -20 dBm ($\sim$ 500 mW electrical power per carrier). The electrical power consumption can be improved down to $<$ 193 mW per carrier (15.5 W total) by reducing the splicing loss between UHNA and SMF fibers \cite{yin2019}, implementing on-chip actuators \cite{liu_2020b} instead of a bulk temperature controller  and using a power-efficient, compact distributed feedback (DFB) laser as CW pump \cite{stern2018, raja2019}. By optimizing the microresonator dispersion design, it is possible to generate 122 comb channels having an optical power > -14 dBm without needing any post-amplification \cite{marin2017}. Similarly, an optimized amplifier utilization with an amplifier for C- and L-band, respectively, would give a comb with an optical power per line $\sim$ 13 dBm in the C- and L- bands mentioned in SI of ref. \cite{marin2017}. 

More importantly, this high power comb source can be shared among multiple servers by adding a hierarchy of passive optical splitters and amplifiers for better power and resource utilization (figure \ref{Fig:figure1}a). The soliton microcomb source distributed among 32 servers provides carriers with $\mathrm{P_{opt}}$ $\sim$ -4 dBm while consuming $\sim$ 2.57 W (1.115 W) electrical power per server by using a state of the art commercial EDFA (on-chip amplifier \cite{juodawlki2011})  making it a highly power-efficient and flexible solution for the data center (cf. SI). More broadly, the flexibility of sharing the comb across many servers, with the fast wavelength selection done on the server itself, means that the overall electrical power efficiency per channel approaches the power efficiency of the EDFAs with the comb power as only a small contributor. This indicates that an optimized shared comb source would consume a comparable electrical power to other multi-wavelength source solutions, including  recent techniques that use a bank of tunable lasers as a multi-wavelength source\cite{gerard2020, sigcomm2020}. Since the wavelength tuning is done on the bank of tunable lasers itself, sharing it between multiple servers would lead to an increased complexity when synchronizing the wavelength switching between the servers, due to the varying time delays between the bank of tunable lasers and the servers. As a matter of fact, one tunable laser bank per server would be more optimal for a time-multiplexed solution \cite{gerard2020} which would require at least 2$\times$32 tunable lasers for switching between 32 servers instead of a single amplified comb chip. Moreover, a comb does not require additional complex algorithms for fast switching and wavelength stabilization, thus offering an appealing division of functionality by leveraging a complex yet highly shared light source. 



In conclusion, ultrafast optical circuit switching is demonstrated using a chip-based microcomb for future power-efficient and low latency data centers. More than twenty individual comb channels in C-band having a power > -20 dBm have been switched at $<$ 550 ps using discrete SOAs.  The optical circuit switching system with 25-GBd NRZ and PAM-4 burst mode data transmission is shown while switching between different comb channels. Further, a PIC containing on-chip SOAs and an AWG is implemented to show sub-ns switching and 25-GBd NRZ transmission. The current demonstration can provide a route for a fully integrated, fast-tunable transceiver providing dense carriers for wavelength switching to meet the power and latency requirements posed by future cloud workloads. 
\\
\begin{footnotesize}
\noindent\textbf{Methods}\\ 
\noindent \textbf{ Switching architecture}: 
The link between two servers is only established via a single wavelength (comb tooth) in  specific time slots ($t_{64}, t_{128}, ...$) as shown in figure \ref{Fig:figure1}a.  The soliton microcombs provide many coherent optical carriers assigned to distinct servers. The switching operation is performed by applying a control signal on the SOAs, e.g., the switching from $t_1$ to $t_2$ data slot is done by applying an on signal to the second SOA and off signals to all other SOAs. A trigger signal from an external reference clock is used to align the switching control and data-encoding units.\\
\noindent \textbf{$\mathbf{Si_3N_4}$ microresoantor}: The $\mathrm{Si_3N_4}$ microresonators are fabricated by using photonic Damascene reflow process enabling waveguides with ultra-smooth sidewalls and linear propagation loss $\sim$ 1 dB\textbackslash m \cite{Pfeiffer2018, liu2020}. The designed microresonators are  over-coupled ($\kappa_{ex} \sim 4 \times \kappa_{0}$) with a FSR of 99.5 GHz and intrinsic linewidth of 15 MHz \cite{Liu2018a}. The waveguides have a  dimension of 1500$\times$900 $\mathrm{nm^2}$. The double inverse nano tapers are used to facilitate the light coupling in- and out- of the chip \cite{liu2018c}.  
\\
\noindent \textbf{Soliton microcomb generation}: A compact fiber laser (Koheras BASIK) is amplified using an EDFA. Then, amplified spontaneous emission (ASE) noise is filtered out using a narrow optical bandpass filter. The $\mathrm{Si_3N_4}$ chip is packaged by splicing the ultrahigh numerical aperture (UHNA) fiber with standard SMF-28 fiber with chip through (fiber-chip-fiber) 15\% coupling efficiency \cite{raja2020}. A single soliton is initiated at an input power of $\sim$ 450 $\mathrm{mW}$ in the bus waveguide by applying a custom designed ramp voltage. The deterministic soliton initiation and backward tuning are controlled via a computer interface. The strong pump line is filtered out using an OADM. Then the soliton is amplified using a low-noise EDFA before de-multiplexing. \\
\noindent \textbf{PIC}: {The InP-based wavelength selector PIC incorporates 23 SOAs of which 4 are used as references. The other 19 SOAs are connected to an 1x32 AWG which acts as a multiplexer and demultiplexer. One of the PIC facets is high-reflection coated so that the light is reflected back through the SOAs and AWG to the input waveguide. The reflective single AWG design reduces the footprint compared to using two AWGs and avoids wavelength misalignment of the AWGs. The wavelength selector PIC was designed using the JEPPIX foundry and fabricated at Fraunhofer HHI. }\\
\noindent \textbf{Switching control unit}: The switching control unit provides the bias currents and electrical switching signals to the SOAs. It also controls the clock and time synchronisation of the switching signals.\\
\noindent \textbf{Funding Information}: This work was supported by a joint research ICES agreement no. MRL Contract No. 2019-034 (01.01.2019 - 31.12.2021). This material is based upon work supported by the Air Force Office of Scientific Research, Air Force Materiel Command, USAF under Award No. FA9550-19-1-0250, and by Swiss National Science Foundation under grant agreement No. 176563 (BRIDGE). This work was supported by funding from the European Union H2020 research and innovation programme PhoMEC under  grant agreement 862528.\\
\noindent \textbf{Acknowledgments}:
The $\mathrm{Si_{3}N_4}$ samples were fabricated and grown in the Center of MicroNanoTechnology (CMi) at EPFL.\\
\noindent \textbf{Author contribution}: A.L. and K.S. designed the $\mathrm{Si_3N_4}$ and InP samples, respectively. J.L. fabricated the $\mathrm{Si_3N_4}$ microresonator. A.S.R., S.L., and M.K. characterized the samples. M.K., X.F., and A.S.R. contributed in breadboard based soliton generator. S.L., A.S.R., and K. S. carried out switching communication experiments. R.B., D.C., I.H., F.K., B.T. and K.J. contributed in the optical transmission system design and optimization. K.S., S.L., and A.S.R.  analysed and processed the data. A.S.R., S.L., H.B., and T.J.K. wrote the manuscript, with the input from others. H.B., T.J.K., and P.C. supervised the project.\\
\noindent \textbf{Data Availability Statement}: The code and data used to produce the plots within this work will be released on the repository \texttt{Zenodo} upon publication of this preprint.
\end{footnotesize}


\cleardoublepage

\bibliographystyle{apsrev4-1}
\bibliography{bibliography}

\end{document}


\title{Supplementary information for Ultrafast optical circuit switching for data centers using integrated soliton microcombs}

\author{A. S. Raja}
\thanks{These authors contributed equally to this work.}
\affiliation{Institute of Physics, Swiss Federal Institute of Technology Lausanne (EPFL), CH-1015 Lausanne, Switzerland}

\author{S. Lange}
\thanks{These authors contributed equally to this work.}
\affiliation{Microsoft Research, 21 Station Road, Cambridge, CB1 2FB, U.K.}

\author{M. Karpov}
\thanks{These authors contributed equally to this work.}
\affiliation{Institute of Physics, Swiss Federal Institute of Technology Lausanne (EPFL), CH-1015 Lausanne, Switzerland}

\author{K. Shi}
\thanks{These authors contributed equally to this work.}
\affiliation{Microsoft Research, 21 Station Road, Cambridge, CB1 2FB, U.K.}

\author{X. Fu}
\affiliation{Institute of Physics, Swiss Federal Institute of Technology Lausanne (EPFL), CH-1015 Lausanne, Switzerland}

\author{R. Behrendt}
\affiliation{Microsoft Research, 21 Station Road, Cambridge, CB1 2FB, U.K.}

\author{D. Cletheroe}
\affiliation{Microsoft Research, 21 Station Road, Cambridge, CB1 2FB, U.K.}

\author{A. Lukashchuk}
\affiliation{Institute of Physics, Swiss Federal Institute of Technology Lausanne (EPFL), CH-1015 Lausanne, Switzerland}

\author{I. Haller}
\affiliation{Microsoft Research, 21 Station Road, Cambridge, CB1 2FB, U.K.}

\author{F. Karinou}
\affiliation{Microsoft Research, 21 Station Road, Cambridge, CB1 2FB, U.K.}

\author{B. Thomsen}
\affiliation{Microsoft Research, 21 Station Road, Cambridge, CB1 2FB, U.K.}

\author{K. Jozwik}
\affiliation{Microsoft Research, 21 Station Road, Cambridge, CB1 2FB, U.K.}

\author{J. Liu}
\affiliation{Institute of Physics, Swiss Federal Institute of Technology Lausanne (EPFL), CH-1015 Lausanne, Switzerland}

\author{P. Costa}
\affiliation{Microsoft Research, 21 Station Road, Cambridge, CB1 2FB, U.K.}

\author{T. J. Kippenberg}
\email[]{tobias.kippenberg@epfl.ch}
\affiliation{Institute of Physics, Swiss Federal Institute of Technology Lausanne (EPFL), CH-1015 Lausanne, Switzerland}

\author{H. Ballani}
\email[]{hitesh.ballani@microsoft.com}
\affiliation{Microsoft Research, 21 Station Road, Cambridge, CB1 2FB, U.K.}

\date{\today}
\maketitle
\section{ Soliton characterization and individual channels switching}

\begin{figure*}[]
\centering
\includegraphics[width=1.\textwidth]{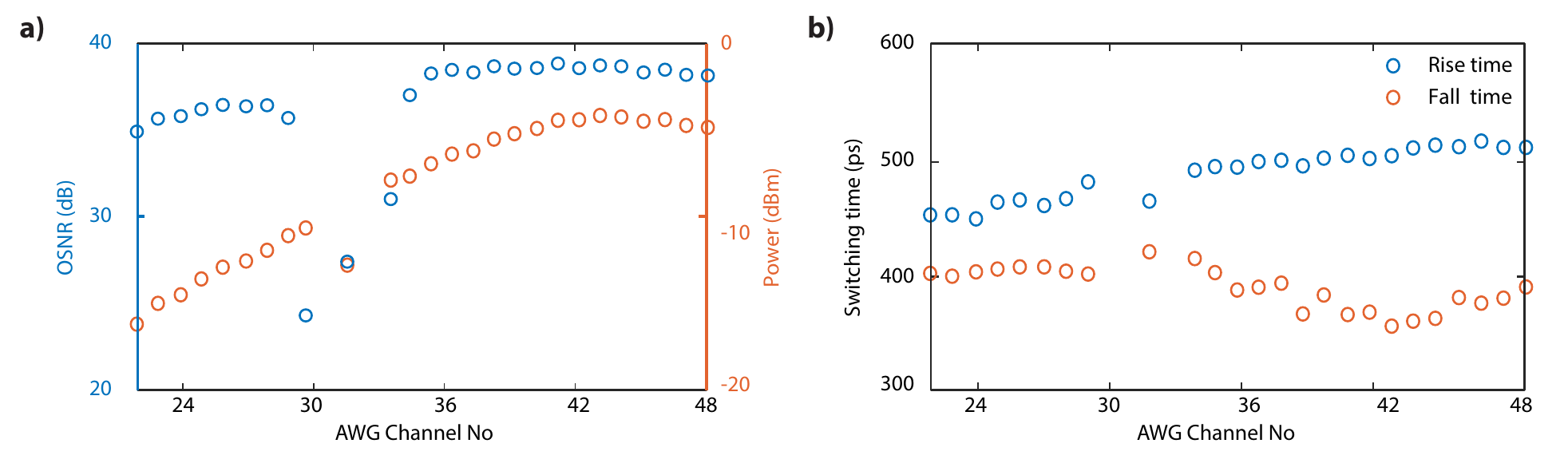}
\caption{\textbf{The power, optical signal to noise ratio (OSNR), and individual channel switching  characterization of soliton microcomb.} \textbf{a)} The optical power of different channels in C-band up to -4 dBm.  The OSNR of two channels around CH 32 (pump mode, $\sim$ 1550 nm) is degraded due to ASE noise. The remaining 23 channels have an OSNR of > 34 dB. \textbf{b)} The 10\% $\times$ 90 \% rise and fall times of around $\sim$ 24 channels lying in the C-band with a maximum switching time of less than 550 ps while using discrete semiconductor optical amplifiers (SOAs). }
\label{Fig:SI_OSNR}
\end{figure*}

Figure \ref{Fig:SI_OSNR} a) shows the absolute power and optical signal to noise ratio (OSNR) of the individual combs lines after post-amplification. Around 25 channels in the C-band have a power of more than -18 dBm and an OSNR of 34 dB.  The four channels around the pump mode centered at 1550 nm (CH 32) have a lower OSNR due to residual ASE noise. This can be avoided either by using a narrower optical bandpass filter or by using the drop-port of the microresonators to couple out the soliton instead of the through port. All these individual comb channels are used to perform the optical circuit switching using discrete SOAs with the setup shown in figure 2a (main text). The 10\% $\times$ 90\% rise and fall times are less than 550 ps. These results indicate that more than 40 channels in L-band can be implemented using the current OCS architecture.

\section{ Four comb channels switching with 20-nm separation}

\begin{figure}[]
\begin{center}
\includegraphics [width=0.5\textwidth]{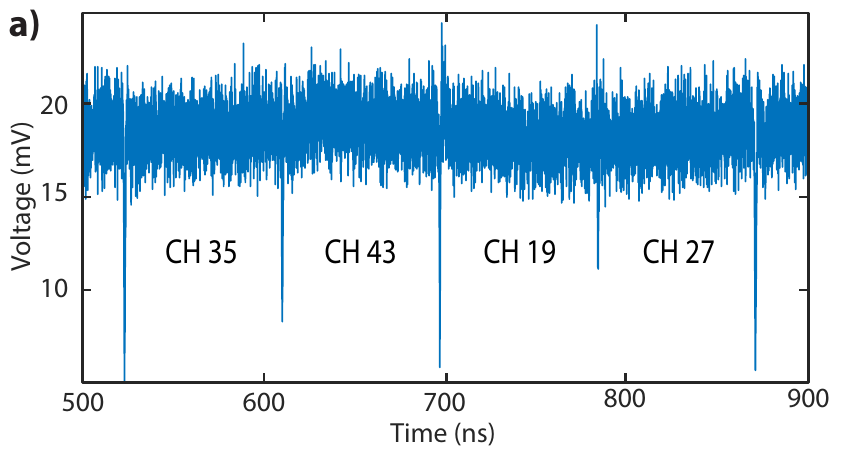}
\caption{\textbf{The fast wavelength switching between four different comb channels spanning 20 nm.} \textbf{a)} The simultaneous switching of four different comb channels using discrete SOAs. The  limitation of the wavelength span in the current system is due to an optical bandpass filter used to reject channels in the next order of free spectral range from C-band AWG.      }
\label{Fig:SI_switching}
\end{center}
\end{figure}

The sub-ns switching between four different comb channels having a maximum separation of more than 20 nm is shown in  figure \ref{Fig:SI_switching}. The variations in comb channel power and SOAs gain are compensated by applying different biasing currents to each SOA. A biasing current of 94 mA, 109 mA, 139 mA, and 134 mA is applied to the SOAs connected  to AWG channels 35, 43, 19, and 27, respectively. 

\section{Soliton microcomb Power consumption}

In this section, a detailed power consumption analysis is carried out to understand the system performance and to provide possible ways to improve it further. The multi-wavelength soliton microcomb is generated using the setup shown in figure 2a (main text).  A compact KOHERAS BASIK fiber laser is used to pump the microresonator, which consumes approximately 1.5 W electrical power when operated at low optical power ($\sim$ 1 mW). A high power BKtel erbium-doped optical amplifier (EDFA) providing gain of up to 34 dB is used mainly to overcome the coupling losses. A single soliton is accessed at a power 1.2 W at the input fiber (450 mW in bus waveguide), while a high power EDFA operated at a  power of 1.8 W compensates the additional 1.5 dB insertion loss (IL) of a narrow band optical bandpass filter. The high power EDFA consumes around 24 W of electrical power. A thermoelectric cooler (TEC) based control loop is implemented to  stabilize the temperature, actuating on the $\mathrm{Si_3N_4}$ chip and consuming $\sim$ 0.5 W of electrical power. After filtering out the pump, the soliton is further amplified using a low noise EDFA which consumes 4 W electrical power. The total electrical power consumed by the multi-wavelength source is 30 W and the electrical power consumed by each carrier is around 500 mW (60 carriers in C- and L-bands suitable for switching and data transmission).
\begin{center}
\begin{tabular}{ |p{5cm}|p{3cm}|  }
\hline
\multicolumn{2}{|c|}{Electrical power consumption (current system)} \\
\hline
Components  & Power consumed \\
\hline
Laser & 1.5 W  \\
TEC & 0.5 W  \\
High power EDFA & 24 W   \\
Low-noise EDFA (post amplification) & 4  \\
Total    & 30  \\
\hline
\end{tabular}
\end{center}
As mentioned in the main text, the main origin of excess power consumption is the lower coupling efficiency into the packaged $\mathrm{Si_3N_4}$ microresonator arising due to splicing loss between the ultra-high numerical aperture (UHNA) and SMF-28 fibers (1 dB per facet). The  coupling efficiency can be improved > 50\% by using an optimized mode converter and by reducing the splicing loss between the UHNA-SMF fibers\cite{yin2019, liu2018c}.  These improvements allow the generation of solitons at an input power of 0.8 W, including 1-dB loss from the optical bandpass filter. In addition, a more compact, power-efficient (electrical power $\sim$ 0.5 W), and on-chip laser can be used to generate the soliton microcomb without requiring lasers with a stabilized module \cite{stern2018, raja2019}. Similarly, a monolithic piezoelectric or an integrated heater element can be used to stabilize the soliton, ensuring long term operation and reducing the power consumption to the range of nW to few-mW (TEC, 0.5 W) \cite{liu_2020b,joshi2016thermally}. These improvements, such as using an on-chip laser, optimized coupling and on-chip actuation, will reduce the power consumption of the system down to 15.5 W (193 mW per carrier). Also, it will give access to more than 80 optical carriers as output coupling will be increased around two times. The  optimization of the microresonators' different design parameters (as shown in supplementary information of ref\cite{marin2017}) such as coupling strength ($\kappa_{ex}$) and second-order dispersion ($\mathrm{D_2}$) eliminates the need for post-amplification of comb lines. This will reduce the power consumption down to 13 W for more than 114 carriers in the C and L bands.  
\begin{center}
\begin{tabular}{ |p{5cm}|p{3cm}|  }
\hline
\multicolumn{2}{|c|}{Electrical power consumption ( $\mathrm{P_{opt.}}$ $\sim $ 13 dBm /line)} \\
\hline
Components  & Power consumed \\
\hline
Laser (DFB) & 0.5 W  \\
High power EDFA & 11 W   \\
Low-noise EDFA (post amplification , C- and L-bands) & 8  \\
High power  EDFA (post amplification , C- and L-bands) & 63 \\
Total    & 82.5  \\
Per server (sharing between 32)    & 2.57  \\
\hline
\end{tabular}
\end{center}

Besides, the comb source can be shared across many servers or transceivers as a parallel multi-wavelength source. As pointed in ref. \cite{marin2017} supplementary information, it is possible to generate the soliton microcomb in C and L-band with $\sim$ 13 dbm  power per carrier while using two additional C- and L-band amplifiers. This particular source can be distributed among 32 servers using a 1$\times$32 splitter with an average insertion loss of around 17 dB. Each server will have a soliton microcomb source with -4 dBm power per carrier for independent wavelength switching. The total electrical power consumed by the whole system and each server is around 82.5 W and 2.57 W respectively. The power efficiency can be improved to 37 W for the whole system and 1.15 W for each server by using the on-chip amplifier\cite{juodawlki2011}. This makes micro-comb based wavelength sources a more flexible and power-efficient solution for optical circuit switching in the data center. 
\cleardoublepage 

\bibliographystyle{apsrev4-1}
\bibliography{bibliography}